\documentclass[man, 12pt, nolmodern, floatsintext, natbib]{apa7}

\usepackage{booktabs}
\usepackage{amsmath}   
\usepackage{amssymb}
\usepackage{graphicx}  
\usepackage{color}     
\usepackage{natbib}    
\usepackage{setspace}
\usepackage{multirow}
\usepackage{mathtools}
\usepackage{array}
\usepackage{tabularx}
\usepackage{scalerel}
\usepackage{framed}
\usepackage{hyperref}
\usepackage[most]{tcolorbox}

\newtcolorbox[
    auto counter
]{procedurebox}[1][]{
    colback=white,
    colframe=black,
    boxrule=0.5pt,
    arc=0pt,
    title={Box~\thetcbcounter:~#1}
}

\renewcommand{\footnoterule}{%
  \kern-3pt \hrule width \textwidth \kern 2.6pt
}

\newcommand{\X}{\mathbf{X}}
\newcommand{\A}{\mathbf{A}}
\newcommand{\B}{\mathbf{B}}
\newcommand{\Set}{\mathbf{S}}

\begin{document}
\title{Covariate Selection for Doubly Robust Double/debiased Machine Learning Estimators for Causal Inference}
\shorttitle{Covariate Selection for DR-DML}
\authorsnames{Muwon Kwon, Peter M. Steiner}
\authorsaffiliations{Department of Human Development and Quantitative Methodology\\ University of Maryland\text{,} College Park}
\authornote{Correspondence concerning this article should be addressed to Muwon Kwon, Department of Human Development and Quantitative Methodology, University of Maryland, College Park, MD 20742, email: mwkwon21@umd.edu.}
\thispagestyle{empty}
\setcounter{page}{0}

\abstract{High-dimensional data create challenges for causal effect estimation because identifying the covariates needed for correct model specification becomes increasingly difficult. Double/debiased machine learning (DML) facilitates the use of machine learning (ML) for causal inference by mitigating regularization and overfitting bias, but comparatively less attention has been given to covariate selection in relation to the double robustness (DR) property possessed by some DML estimators. In particular, ML-based covariate selection may result in differential covariate selection or in misspecification of both models, thereby limiting the practical utility of the DR property. To address these issues, we propose using the union of the covariates selected by the propensity score (PS) and outcome ML models to re-estimate both models. Simulation results show that using the union consistently reduces more confounding bias than using separate selected covariate sets. The results also show that ML-based estimation does not uniformly outperform conventional DR estimation, even under conditions favorable to the Lasso, and that post-Lasso reduces more confounding bias than standard Lasso. These findings demonstrate that successful use of ML for causal inference depends not only on the ML algorithm but also on how the information obtained through covariate selection is incorporated into causal effect estimation.

\vspace{7mm}
}

\maketitle

\section{Introduction}
The consistent estimation of causal effects from observational studies requires a set of observed covariates that removes all confounding bias \citep[unconfoundedness assumption; ][]{rosenbaum1983central}. In high-dimensional settings where the number of covariates is large relative to the sample size, it becomes demanding to select such a covariate set and correctly specify the functional form of the outcome model. To address the difficulty with high-dimensional data, researchers have been relying on machine learning (ML) methods such as the least absolute shrinkage and selection operator (Lasso; \citealp{tibshirani1996regression}) because they automate covariate selection and provide a systematic framework for learning the conditional expectation function from the data \citep{belloni2014inference}.
However, the common practice of applying ML only to the outcome model to estimate effects of interest faces two sources of bias: (1) regularization bias and (2) overfitting bias, which prevent the ML-based effect estimators from achieving $\sqrt{n}$-consistency, where $n$ is the sample size \citep{10.1111/ectj.12097}. Regularization constrains model complexity by imposing penalties on the estimation procedure to prevent a model from fitting the training data extremely well, thereby improving out-of-sample prediction.\footnote{For example, the penalty term of the Lasso reduces the complexity of the model by shrinking the overall magnitude of the estimated coefficients towards zero, with some coefficients being shrunk exactly to zero. Also, random forest regularizes the model by limiting the correlation among individual trees through bootstrap sampling and random feature selection, and then averaging their predictions to reduce variance and improve generalization.} Consequently, the regularization keeps the variance of the estimator from inflating but also necessarily induces bias in the estimator. Overfitting bias occurs when the same sample used to estimate the treatment selection and outcome models is also used to estimate the causal effect.\footnote{In the ML literature, the treatment selection and outcome models are commonly referred to as nuisance models, whereas the causal effect is referred to as the target parameter.} When the two models are estimated using ML methods, they may overfit the estimation sample by capturing not only the underlying systematic relationships but also idiosyncratic noise specific to the sample. If the same observations are subsequently used to estimate the causal effect, this sample-specific overfitting may influence the estimated causal effect, resulting in biased estimates.

To address issues due to regularization and overfitting with high-dimensional data and to obtain $\sqrt{n}$-consistent effect estimators, \citet{10.1111/ectj.12097} proposed double/debiased ML (DML). DML mitigates the impact of the regularization bias by constructing estimators that are locally insensitive to small estimation errors in the treatment selection and outcome models, a property known as Neyman orthogonality \citep{Neyman1959}. This local insensitivity limits the impact of the regularization bias on the causal effect estimators when the estimation errors in the two models converge to zero sufficiently fast.
To mitigate overfitting bias, DML employs sample splitting, ensuring that the treatment selection and outcome models are estimated using one subset of the data, while the causal effect is estimated using a separate, non-overlapping subset. This separation of samples reduces the extent to which sample-specific overfitting in the two estimated models influences the causal effect estimate.

Because DML enables $\sqrt{n}$-consistent estimation of the causal effect despite estimation errors in the estimated treatment selection and outcome models, it has become an increasingly popular framework for estimating causal effects in high-dimensional settings \citep{chernozhukov2024applied, valentin2025double, hunermund2023double}. In addition to this primary and well-established convergence-rate property shared by \textit{all} DML estimators, there is another robustness property exhibited by \textit{some} DML estimators that has not yet been explicitly emphasized in the DML literature: the double robustness (DR; also referred to as doubly robust) property \citep{valentin2025double, chernozhukov2024applied}. The DR property ensures consistency of the estimator if either the treatment selection model or the outcome model is correctly specified \citep{robins1994estimation, robins1995semiparametric, glynn2010introduction, scharfstein1999adjusting, tsiatis2006semiparametric, robins2001comment}. 
That is, the DR property offers protection against model misspecification because it provides two opportunities to remove confounding bias, either by correctly specifying the treatment selection model or the outcome model relative to the joint covariate set that meets unconfoundedness. Correct specification of these models involves two aspects: the selection of covariates to be controlled for and the functional form of the models. This article focuses exclusively on identifying appropriate covariate sets for doubly robust DML (DR-DML) estimators. Accordingly, we do not address the specification of the functional forms of the two models, assuming that in high-dimensional settings, their functional forms can be adequately estimated using ML methods.

Although the DR property provides protection against model misspecification, current implementation of DML may limit the practical utility of the DR property because of two issues arising from incorrect covariate selection: (1) differential covariate selection \citep{steiner2024robust} and (2) misspecification of both the treatment selection and outcome models. Differential covariate selection refers to the situation in which the treatment selection and outcome models of DR estimators adjust for different sets of covariates. Under differential covariate selection, the DR property may not hold even when one or both models would be correctly specified on their own because the seemingly correctly specified model is always kept blind to any amplification-bias or collider-bias-inducing misspecifications of the respective other model. When ML methods are used to estimate the treatment selection and outcome models in high-dimensional settings, differential covariate selection may commonly arise even when both models are specified using the same initial set of covariates because ML methods rarely select exactly the same subset of covariates for both models. Moreover, in practice it is much more likely that both models are misspecified, that is, each model fails to remove the entire confounding bias. In this case the DR property no longer provides complete protection against model misspecification (though using a DR estimator might often nonetheless be beneficial even though not all bias is removed). This may occur even when the covariate sets initially included in the treatment selection and outcome ML models contain all covariates needed for approximately correct specification of at least one model. Because the ML performance for covariate selection depends on factors such as sample size and the choice of tuning parameters, ML methods may fail to select all covariates necessary for correct model specification from the initial covariates that meet unconfoundedness.


To avoid or at least mitigate bias arising from incorrect covariate selection with DR-DML estimators, we propose using the union of the two model-specific covariate sets to re-estimate the treatment selection and outcome ML models. The union of the covariate sets is used in the double selection approach \citep{belloni2014inference} to ensure that important covariates for both models are included in the outcome model, thereby mitigating omitted-variable bias when estimating the causal effect. However, the primary emphasis of the double selection approach is on applying Lasso to both the treatment selection and outcome models, rather than on exploiting the role of the union itself in model specification. Consequently, the use of the union in double selection is limited to the particular model specification and data-generating process (DGP) considered. Specifically, the union is used only to specify the final outcome model under a DGP that assumes a constant treatment effect. 

Beyond this limited use of the union in the double selection, we fully leverage the distinct advantages of DR-DML estimators and the union set of selected covariates for causal effect estimation. In DR-DML estimators, after covariate sets are selected separately for the treatment selection and outcome models using their respective ML methods, we re-estimate both models using the union of the selected covariate sets. Using the union eliminates differential covariate selection, thereby allowing each model to account for collider-bias-inducing or bias-amplifying misspecifications arising from covariates selected by the other model. In addition, the union increases the likelihood that at least one model includes the covariates necessary for approximately correct specification, thereby helping preserve the DR property. In particular, using the union may allow the DR property to hold even when both models are misspecified based on their respective selected covariate sets. Although neither selected covariate set may correctly specify its corresponding model, the union of the two sets may be sufficient to correctly specify at least one of them. re-estimating both models using the union therefore provides an additional opportunity for at least one model to be correctly specified, thereby preserving the DR property.

In this study, we use the Lasso as the ML method because it is one of the most widely used approaches for automated covariate selection \citep{belloni2011inference, belloni2012sparse, belloni2014inference} and has played a foundational role in the theoretical development of DML \citep{chernozhukov2024applied}. By focusing on the Lasso, we explicitly distinguish the respective advantages of DR-DML estimation and the union of selected covariates and examine how these advantages can be combined for ATE estimation. Our primary emphasis in this study is on reducing confounding bias in finite samples rather than on the asymptotic $\sqrt{n}$-consistency of the estimators, although the proposed approach is designed to preserve the $\sqrt{n}$-consistency of DR-DML estimators. 

Beyond demonstrating the benefits of DR-DML estimation and the use of the union set, this study also provides insight into when ML-based covariate selection is beneficial and when it may be unnecessary or even undesirable. As the use of ML methods has become increasingly widespread, there is a risk of implicitly assuming that ML-based procedures necessarily yield more reliable causal effect estimates than conventional methods. Our simulation results help clarify the conditions under which ML-based covariate selection provides practical advantages and those under which conventional estimation remains preferable. We further examine the advantage of post-Lasso estimation, which mitigates the shrinkage induced by the Lasso by re-estimating the model using the selected covariates without penalization. Although the advantages of post-Lasso over Lasso have been well established \citep{10.3150/11-BEJ410}, the two approaches are often used or discussed interchangeably in applied research. In the simulation study, we compare standard Lasso and post-Lasso within the same DR-DML framework and show that post-Lasso yields a greater reduction in confounding bias than standard Lasso.

\section{Double/debiased Machine Learning}
\citet{10.1111/ectj.12097} proposed DML to address two major sources of bias that arise when ML methods are used for causal effect estimation: regularization bias and overfitting bias. This section briefly describes why these biases arise and how DML addresses them. In our discussions, we particularly focus on regularization bias because the DML strategy used to avoid this bias is more directly related to the DR property of DML estimators---our primary focus. Readers interested in more detailed explanations of regularization and overfitting bias, including technical derivations and formal results, are referred to \cite{10.1111/ectj.12097, chernozhukov2024applied}. 

To formalize ideas, we consider the estimation of the average treatment effect (ATE) under a DGP with heterogeneous treatment effects:
\begin{equation} \label{equ:IRM}
\begin{aligned}
    Y & = g(Z, \X) + U, \hspace{1em} \text{E}[U | \X, Z] = 0, \\
    Z & = m(\X) + V, \hspace{1.8em} \text{E}[V | \X] = 0,
\end{aligned}
\end{equation}
where $Y$ is the outcome, $Z$ is a binary treatment indicator with $Z = 1$ denoting treatment and $Z = 0$ denoting control condition, and $\X$ is a set of observed confounders that satisfies the unconfoundedness assumption.
$U$ and $V$ are additive disturbances. The confounders $\X$ influence both treatment assignment and the outcome through two respective models: the propensity score (PS) function $m(\X) = \text{P}(Z = 1 \mid \X)$ for treatment assignment and the potentially nonlinear outcome function $g(Z, \X) = \text{E}[Y \mid Z, \X]$ for the outcome. The target parameter of interest is the ATE:
\begin{equation} \label{equ:theta_g}
\begin{aligned}
\theta = \text{E}[g(1, \X) - g(0, \X)],
\end{aligned}
\end{equation}
that is, the expected difference in the predicted treatment and control outcomes.

\subsection{Regularization and Overfitting Bias in ML-based Causal Inference}
\subsubsection{Regularization Bias}
When estimating the ATE via ML with high-dimensional data, a naive approach is to estimate the ATE $\theta$ using only the outcome model in Equation \ref{equ:IRM} while ignoring the selection model. However, while applying ML to the outcome model offers flexible and accurate predictions of the outcome, it does not imply that one obtains an unbiased estimator for the ATE. Suppose the outcome model $g(Z, \X)$ in Equation \ref{equ:IRM} is estimated using a regularized ML method. Then, the regression estimator can be decomposed as follows: 
\begin{equation} \label{equ:err_outcome_model}
\begin{aligned}
\hat{g}(z, \X) = g(z, \X) + \delta_g(z, \X),
\end{aligned}
\end{equation}
where $\delta_g(z,\X)$ denotes the estimation error. When a regularized ML method is used, this estimation error has non-zero mean because regularization results in a systematic error component. Regularization controls model complexity by imposing penalties, restrictions, or other constraints on the estimation procedure to obtain a more parsimonious model that avoids overfitting and, therefore, results in improved predictive performance on out-of-sample observations.
For example, the Lasso regularizes a regression model by imposing a penalty that shrinks estimated coefficients toward zero and sets some coefficients exactly to zero. Random forests control model complexity and reduce prediction variance through averaging across trees constructed from bootstrap samples and randomly sampled subsets of predictors. Although regularization prevents the variance of an ML estimator from becoming excessively large, it introduces systematic bias into the estimator. This bias arising from regularization is referred to as \textit{regularization bias}. 
Consequently, the estimator $\hat{g}(z,\X)$ is biased for $g(z,\X)$, with bias given by $\text{E}[\delta_g(z,\X)]$. 

The impact of regularization bias in the regression estimator $\hat{g}(z,\X)$ on the ML-based ATE estimator $\hat{\theta}_{\text{ML}}$ can be illustrated analytically using the formulation in \citet{10.1111/ectj.12097}.\footnote{The DML literature \citep{10.1111/ectj.12097, chernozhukov2024applied} discusses regularization bias in the context of semi-parametric score functions. Here, we illustrate the same issue using the simpler plug-in estimator.} Suppose that the ATE is estimated using $\hat{g}(z,\X)$. The resulting plug-in estimator is
\begin{equation} \label{equ:ML estimator}
\begin{aligned}
\hat{\theta}_{\text{ML}}
&= \mathbb{E}_n[\hat{g}(1,\X)-\hat{g}(0,\X)] \\
&= \mathbb{E}_n[g(1, \X) + \delta_g(1, \X)] - \mathbb{E}_n[g(0, \X) + \delta_g(0, \X)],
\end{aligned}
\end{equation}
where the last equality follows from the decomposition in Equation \ref{equ:err_outcome_model}. Then the overall estimation error of the ML estimator $\hat{\theta}_{\text{ML}}$ can be decomposed as follows:
\begin{equation} \label{equ:deviation_ml}
\begin{aligned}
\hat{\theta}_{\text{ML}} - \theta 
&= \mathbb{E}_n[g(1, \X) + \delta_g(1, \X)] - \mathbb{E}_n[g(0, \X) + \delta_g(0, \X)] - \text{E}[g(1, \X) - g(0, \X)] \\
&= \underbrace{\mathbb{E}_n[g(1, \X) - g(0, \X)] - \text{E}[g(1, \X) - g(0, \X)]}_{\text{sampling error}} + \underbrace{\mathbb{E}_n[\delta_g(1, \X) - \delta_g(0, \X)]}_{\text{first order estimation error}}.
\end{aligned}
\end{equation}
The first term captures sampling error arising from replacing the population expectation with its empirical counterpart. The sampling error term has mean zero and typically converges to zero at the rate $n^{-1/2}$, which is compatible with $\sqrt{n}$-consistent estimation. The second term captures the estimation error in the outcome regression. Recall that $\delta_g(z,\X)$, for $z = 0, 1$, contains a systematic component induced by regularization and therefore has a non-zero mean. Moreover, $\delta_g(z,\X)$ converges to zero at a rate slower than $n^{-1/2}$. Because this error enters the plug-in ATE estimator at first order—that is, linearly—its slower convergence rate directly carries over to the ATE estimator, thereby preventing $\sqrt{n}$-consistent estimation of the ATE. Taking the expectation of Equation~\ref{equ:deviation_ml} eliminates the mean-zero sampling-error term, yielding the bias of the ML estimator:
\begin{equation} \label{equ:bias_ml}
\begin{aligned}
\text{E}[\hat{\theta}_{\text{ML}}]-\theta
&=
\text{E}[\delta_g(1,\X)-\delta_g(0,\X)].
\end{aligned}
\end{equation}
Equation~\ref{equ:bias_ml} shows that the bias of the ML estimator is directly determined by first order regularization bias of the outcome regression estimator.

\subsubsection{Overfitting Bias}
Overfitting bias occurs when the same sample used to estimate the PS and outcome models, $g(Z, \X)$ and $m(\X)$, is also used to estimate the causal effect $\theta$. When the two models are estimated using ML methods, they tend to overfit the estimation sample by capturing not only the underlying systematic relationships but also the sample-specific noise. If the same data are subsequently used to estimate the causal effect, this sample-specific overfitting potentially biases the causal effect estimate.

\subsection{Mitigating Regularization Bias}
DML mitigates the impact of regularization bias and other first order estimation errors by constructing estimators that incorporate not only the outcome model but also the PS model. For the DGP specified in Equation \ref{equ:IRM}, the corresponding DML estimator is constructed as follows\footnote{Under a DGP that assumes a constant treatment effect, DML estimators are constructed differently. For details, see \citep{chernozhukov2024applied, 10.1111/ectj.12097}.} \citep{belloni2014inference, 10.1111/ectj.12097}:
\begin{equation} \label{equ:dml estimator}
\begin{aligned}
\hat{\theta}_{\text{DR-DML}} 
&= \mathbb{E}_n\Bigg[\hat{g}(1, \X) - \hat{g}(0, \X) + \cfrac{Z(Y - \hat{g}(1, \X))}{\hat{m}(\X)} - \cfrac{(1 - Z)(Y - \hat{g}(0, \X))}{1- \hat{m}(\X)}\Bigg].
\end{aligned}
\end{equation}
Note that we use the acronym DR-DML to denote this estimator because it takes the well-established augmented inverse probability weighting (AIPW; \citealp{robins1994estimation, robins2001comment, scharfstein1999adjusting}) form, which possesses the DR property discussed in detail in the following sections. Accordingly, this estimator may be referred to as either the DML estimator or the DR-DML estimator. In this section, however, we use the term DML estimator to focus on its role in mitigating regularization and overfitting bias, and reserve the term DR-DML estimator for the subsequent sections after introducing its DR property.

While the ML estimator in Equation \ref{equ:ML estimator} is constructed using only the outcome model, the DML estimator in Equation \ref{equ:dml estimator} relies on both the PS and outcome models. Analogous to the decomposition in Equation \ref{equ:err_outcome_model}, an ML-based PS estimator, $\hat{m}(\X)$, can also be decomposed:
\begin{equation} \label{equ:err_ps_model}
\begin{aligned}
\hat{m}(\X) &= m(\X) + \delta_m(\X),
\end{aligned}
\end{equation}
where $\delta_m(\X)$ denotes the estimation error in the PS estimator. As before, this estimation error has a non-zero mean due to a systematic component induced by regularization. Thus, $\hat{m}(\X)$ is biased for $m(\X)$, with bias given by $\text{E}[\delta_m(\X)]$. Substituting the two decompositions of the PS and outcome estimators from Equations \ref{equ:err_outcome_model} and \ref{equ:err_ps_model} into the ATE estimator in Equation \ref{equ:dml estimator} yields
\begin{equation}
\begin{aligned}
\hat{\theta}_{\text{DR-DML}} 
&= \mathbb{E}_n\Bigg[
g(1,\X) + \delta_g(1,\X)
- g(0,\X) - \delta_g(0,\X) \\
&\quad\quad\quad
+ \frac{Z\{Y-g(1,\X)-\delta_g(1,\X)\}}
{m(\X)+\delta_m(\X)} 
- \frac{(1-Z)\{Y-g(0,\X)-\delta_g(0,\X)\}}
{1-m(\X)-\delta_m(\X)}
\Bigg].
\end{aligned}
\end{equation}
Then the bias of the ATE estimator $\hat{\theta}_{\text{DR-DML}}$ is given by (see the Appendix for details):
\begin{equation} \label{equ:bias_dml}
\begin{aligned}
\text{E}[\hat{\theta}_{\text{DR-DML}}] - \theta 
&=
\text{E}\left[
\frac{\delta_g(1,\X)\delta_m(\X)}
{m(\X)+\delta_m(\X)}
+
\frac{\delta_g(0,\X)\delta_m(\X)}
{1-m(\X)-\delta_m(\X)}
\right].
\end{aligned}
\end{equation}
The equation highlights that the bias of the DML estimator depends on the products of the estimation errors in the PS and outcome estimators. Because of the multiplication of independent errors, their joint bias contribution is of second order and, therefore, converges to zero faster than either first order estimation error alone. Consequently, even when each estimator individually converges at a rate slower than $n^{-1/2}$, the DML estimator can remain $\sqrt{n}$-consistent as long as the product of the two estimation errors converges to zero at the rate $n^{-1/2}$ or faster. For example, even if the estimation errors in both the PS and outcome regression estimators converge at the rate $n^{-1/4}$, their product converges at the rate $n^{-1/4}\times n^{-1/4}=n^{-1/2}$. This insensitivity to estimation errors in the two models arises from Neyman orthogonality \citep{Neyman1959} of the underlying score function. For detailed treatments of Neyman orthogonality, including formal derivations and mathematical proofs, see \citet{10.1111/ectj.12097, chernozhukov2024applied}.

\subsection{Removing Overfitting Bias}
To remove overfitting bias, DML employs sample splitting, ensuring that the PS and outcome models are estimated using one subset of the data, while the causal effect is estimated using a separate, non-overlapping subset. This separation of samples reduces the extent to which sample-specific overfitting in the two estimated models influences the causal effect estimate. While sample splitting prevents overfitting bias in the estimators, its direct application can lead to a substantial loss of efficiency because the causal effect is estimated using only a subset of the available data. This inefficiency can be addressed through cross-fitting, in which the roles of the separate samples are reversed to obtain additional estimates of the causal effect. Averaging these estimates allows all observations to contribute to the final estimate, thereby recovering much of the efficiency lost through sample splitting \citep{10.1111/ectj.12097}.

\section{Double Robustness of DML Estimators}
As briefly mentioned above, the DML estimator in Equation \ref{equ:dml estimator} is not only $\sqrt{n}$-consistent, a property arising from Neyman orthogonality that makes the estimator insensitive to small estimation errors in both the PS and outcome models, but also possesses the DR property. The DR property ensures consistency of the estimator if either the PS model or the outcome model is correctly specified with regard to a joint covariate set $\X$ that establishes unconfoundedness \citep{robins1994estimation, robins1995semiparametric, glynn2010introduction, scharfstein1999adjusting, tsiatis2006semiparametric, robins2001comment, steiner2024robust}.\footnote{Neyman orthogonality can be viewed as a form of robustness because it makes the estimator insensitive to small perturbations in the estimated PS and outcome models. However, to follow the terminology commonly used in the DML literature \citep{10.1111/ectj.12097, chernozhukov2024applied}, we distinguish the local insensitivity property from the DR property throughout this article.} That is, the DR property offers protection against model misspecification by providing two opportunities to remove the entire confounding bias, either by correctly specifying the PS model or the outcome model. Because of this robustness to model misspecification, DR estimation has been widely used in causal inference \citep{bang2005doubly, glynn2010introduction, imbens2009recent, schafer2008average}. 
The DR property of the DML estimator in Equation \ref{equ:dml estimator} can be expressed as follows:
\begin{equation}
\widehat{\theta}_{\text{DR-DML}}
\xrightarrow{p}
\theta
\quad\text{if}\quad
\widehat{g}(z, \mathbf{R}_g) \xrightarrow{p} g(z, \X)
\quad\text{or}\quad
\widehat{m}(\mathbf{R}_m) \xrightarrow{p} m(\X),
\end{equation}
where $\mathbf{R}_g$ and $\mathbf{R}_m$ denote the covariate sets used to specify the outcome and PS models, respectively. Thus, consistency of the DR-DML estimator requires that at least one of the two models be correctly specified with regard to $\X$. Proofs of this result can be found in \citet{glynn2010introduction} or \citet{tsiatis2006semiparametric}. Note that the correct specification of a model involves two aspects: the correct selection of covariates $\mathbf{R}$ to be controlled for and the correct specification of the functional relation $\widehat{g}(.)$ or $\widehat{m}(.)$, respectively. Thus, a correctly specified model must control for a set of covariates that satisfies unconfoundedness and correctly capture the functional relationship between the dependent variable and the covariates and address bias-inducing misspecification arising from the misspecified model. In this study, we focus on the DR property within the DML framework in high-dimensional settings, with particular emphasis on covariate selection. Accordingly, we do not consider potential misspecification of the functional forms of the PS and outcome models. 

Importantly, not all DML estimators possess the DR property. Neyman orthogonality and DR are distinct properties: Neyman orthogonality does not, by itself, imply DR, nor does DR necessarily imply Neyman orthogonality. A DML estimator has the DR property only when its underlying score function possesses the DR property. For example, DML estimators for DGPs with constant treatment effects employ score functions that satisfy Neyman orthogonality but do not have the DR property \citep{10.1111/ectj.12097}. This distinction is important because the terms Neyman orthogonality and the DR property are sometimes used interchangeably in discussions of DML, which can lead to the misconception that all DML estimators have the DR property. For a detailed discussion of the DR property of the DML estimators, see \citet{valentin2025double, chernozhukov2024applied}. 


\subsection{Challenges of Covariate Selection in ML-Based DR Estimation}
Although the DR property of the DML estimator ensures consistency if either the PS model or the outcome model is correctly specified, identifying the covariates needed for correct specification can still be challenging in practice. In particular, two major issues may potentially arise from incorrect covariate selection: (1) differential covariate selection \citep{steiner2024robust}, which refers to the situation in which the PS and outcome models adjust for different sets of covariates, and (2) misspecification of both models, which refers to the situation in which each covariate set selected for the respective model fails to contain all covariates necessary for correct model specification. These two problems are commonly encountered in practice, regardless of whether ML methods are used. When ML methods are not used, these problems are primarily related to the quality and validity of the subject-matter knowledge researchers have about the underlying true DGP. However, when ML methods are used, which is the focus of this article, these problems may arise not only from limitations in subject-matter knowledge but also from the covariate selection procedures employed by the ML methods.

To illustrate these challenges, suppose that we estimate the ATE under the DGP in Equation \ref{equ:IRM} using double post-Lasso \citep{belloni2014inference}, which applies post-Lasso \citep{10.3150/11-BEJ410} separately to the PS and outcome models. Post-Lasso builds on the Lasso by separating covariate selection from subsequent model estimation. The Lasso penalty shrinks estimated coefficients toward zero, setting some exactly to zero and thereby selecting covariates. Post-Lasso removes this shrinkage by re-estimating the models without penalization using the covariates selected by the initial Lasso, with logistic regression for the PS model and ordinary least squares (OLS) for the outcome model. Suppose that, based on reasonable subject-matter theory, we initially include the covariate set $\X$ that satisfies unconfoundedness in both the PS and outcome regression Lasso models.
The procedure for estimating the ATE is then implemented as follows:
\begin{procedurebox}[Estimation Procedure of Double Post-Lasso]
\label{box:doublelasso}
\begin{enumerate}
    \item Covariate selection using the Lasso.
    \begin{itemize}
        \item For the PS model $m(\X)$.
        \begin{enumerate}
            \item Fit a logistic Lasso regression of $Z$ on $\X$.
            \item Identify the selected covariates $\X_{\text{PS}} \subset \X$ with non-zero coefficients.
        \end{enumerate}
        
        \item For the outcome model $g(z, \X)$ for $z = 0, 1$.
        \begin{enumerate}
            \item Fit a Lasso regression of $Y_z$ on $\X$.
            \item Identify the selected covariates $\X_{Y_z} \subset \X$ with non-zero coefficients.
        \end{enumerate}
    \end{itemize}

    \item Re-estimation using the selected covariate set.
    \begin{itemize}
        \item For the PS model $m(\X)$.
        \begin{enumerate}
            \item Fit a logistic regression of $Z$ on $\X_{\text{PS}}$.
            \item Predict the PSs using the selected set: $\hat{Z} = \hat{m}(\X_{\text{PS}})$. 
        \end{enumerate}

        \item For the outcome model $g(z, \X)$ for $z = 0, 1$.
        \begin{enumerate}
            \item Fit an OLS regression of $Y_z$ on $\X_{Y_z}$.
            \item Predict the outcomes using the selected set: $\hat{Y}_z = \hat{g}(z, \X_{Y_z})$. 
        \end{enumerate}
    \end{itemize}

    \item Estimation of the ATE by plugging the predicted values $\hat{m}(\X_{\text{PS}})$ and $\hat{g}(z, \X_{Y_z})$ into Equation \ref{equ:dml estimator}.
\end{enumerate}
\end{procedurebox}
\noindent In the procedure described in \hyperref[box:doublelasso]{Box 1}, the re-estimation step represents the key distinction of the post-Lasso approach. If the PSs and outcomes are predicted directly from the fitted Lasso models without this re-estimation step, the procedure corresponds to the double Lasso approach \citep{belloni2014inference}.

\subsubsection{Differential Covariate Selection}
In the double post-Lasso procedure described in \hyperref[box:doublelasso]{Box 1}, if the PS and outcome models end up controlling for different covariate sets (i.e., $\X_{\text{PS}} \neq \X_{Y_z}$), the DR property may not hold even if one or both models are correctly specified. This occurs because the seemingly correctly specified model is always kept blind to any collider-bias-inducing or bias-amplifying misspecifications of the respective other model \citep{steiner2024robust}. For example, consider a DGP that differs from the one in Equation \ref{equ:IRM}, in which the treatment indicator is affected by both confounders and instrumental variables. In this DGP, suppose that the PS model is correctly specified by controlling for all confounders required to satisfy unconfoundedness, whereas the outcome model is misspecified because it omits some confounders while including instrumental variables that are not included in the PS model. Then, the PS estimator is unbiased, whereas the outcome regression estimator is biased because of the omitted confounders, with the bias further amplified by conditioning on the instrumental variables \citep{steiner2016mechanics}. In this case, although the PS model is correctly specified, the DR property does not hold, resulting in biased effect estimates. This occurs because the PS model fails to account for the additional bias that is amplified when the outcome model conditions on the instrumental variables. Consequently, although the PS model successfully removes the bias induced by the confounders, it remains blind to the bias-amplifying misspecification in the outcome model. For further examples and detailed explanations of differential covariate selection, see \citet{steiner2024robust}.

Although the PS and outcome regression Lasso models initially include the common covariate set $\X$, the two models may \textit{end up} adjusting for different sets of covariates because ML methods rarely select exactly the same subset of covariates for both models due to, for instance, differences in the predictive strength of covariates across the two models.\footnote{Differential covariate selection is more likely to occur when the initially included covariate sets differ between the PS and outcome regression Lasso models. Such a specification can be desirable because using a smaller subset of covariates in the PS model can help maintain covariate overlap (i.e., the positivity assumption), whereas the outcome model should include all covariates predictive of the outcome to improve the efficiency of the treatment effect estimator \citep{cho2024variable, de2011covariate, glynn2010introduction, koch2018covariate, lunceford2004stratification, vansteelandt2012model, witte2019covariate}.}

\subsubsection{Misspecification of Both PS and Outcome Models}
While differential covariate selection can be problematic even when one or both of the PS and outcome models are correctly specified, correct specification of at least one model is not always guaranteed in practice. In other words, both models may be misspecified, in which case the DR property no longer provides protection against model misspecification. In the double post-Lasso procedure described in \hyperref[box:doublelasso]{Box 1}, even when the initially included covariate set $\X$ is sufficient for approximately correct specification of at least one model, the selected covariate sets, $\X_{\text{PS}}$ and $\X_{Y_z}$, may fail to retain all covariates necessary for such specification of their respective models. Whether this occurs depends on the performance of the corresponding Lasso models, which may be affected by several factors, including sample size, the approximate sparsity assumption, and the choice of the penalty parameter.\footnote{Also, because the true DGP is rarely known, the initially included covariate set $\X$ for both models may not be sufficient to approximately specify their respective models. In such cases, the final covariate sets selected by the Lasso ($\X_{\text{PS}}$ and $\X_{Y_z}$) are even less likely to be sufficient for approximate specification.}

\section{Using Union of Covariate Sets for Doubly Robust DML Estimators}
Given that these two issues arising from incorrect covariate selection may prevent the DR property from holding in practice, we address both to increase the likelihood that the DR property is preserved in DR-DML estimation. In particular, we propose using the \textit{union} of the covariate sets selected by the PS and outcome regression ML models to re-estimate both models. The union set is used in the double selection approach \citep{belloni2014inference}, in which the primary strategy is to apply Lasso separately to the PS and outcome models. 
The covariates selected by the two Lasso models are combined into a union set, which is used to re-estimate the final outcome model. In this way, covariates selected as important for either the PS or outcome model are retained in the final regression, thereby reducing the risk of omitted-variable bias in treatment effect estimation.

However, the role of the union in double selection is limited to DGPs with constant treatment effects. The double selection approach does not use the union under DGPs with heterogeneous treatment effects. Under heterogeneous treatment effects, ATE estimation instead follows the procedure described in \hyperref[box:doublelasso]{Box 1}, in which the PS and outcome models are re-estimated using their separately selected covariate sets. Consequently, this procedure remains susceptible to the two problems arising from incorrect covariate selection discussed above. This distinction highlights that, in double selection, the union primarily serves as a mechanism for integrating the covariate selection results from the two Lasso models under DGPs with constant treatment effects, rather than as a general strategy for re-estimating both the PS and outcome models to increase the likelihood that the DR property holds.

Our proposal is to distinguish the respective advantages of DR-DML estimators and the union set and to combine them explicitly in ATE estimation. Specifically, we use the union set to re-estimate both the PS and outcome models, with the goal of addressing the two problems arising from incorrect covariate selection and, in turn, increasing the likelihood that the DR property holds. Suppose again that we estimate the ATE under the DGP in Equation \ref{equ:IRM} using the double post-Lasso approach, but now incorporating the union set. As before, we initially include the covariate set $\X$ that meets unconfoundedness in both the PS and outcome regression Lasso models. The estimation procedure using the union set of covariates is then implemented as follows:
\begin{procedurebox}[Estimation Procedure of Double Post-Lasso Using the Union Set]
\label{box:doublelassounion}
\begin{enumerate}
    \item Covariate selection using the Lasso.
    \begin{itemize}
        \item For the PS model $m(\X)$.
        \begin{enumerate}
            \item Fit a logistic Lasso regression of $Z$ on $\X$.
            \item Identify the selected covariates $\X_{\text{PS}} \subset \X$ with non-zero coefficients.
        \end{enumerate}
        
        \item For the outcome model $g(z, \X)$ for $z = 0, 1$.
        \begin{enumerate}
            \item Fit a Lasso regression of $Y_z$ on $\X$.
            \item Identify the selected covariates $\X_{Y_z} \subset \X$ with non-zero coefficients.
        \end{enumerate}
    \end{itemize}

    \item Construction of the union set: $\X_{\text{Union}} = \X_{\text{PS}} \cup \X_{Y_0} \cup \X_{Y_1}$.

    \item Re-estimation using the union set. 
    \begin{itemize}
        \item For the PS model $m(\X)$.
        \begin{enumerate}
            \item Fit a logistic regression of $Z$ on $\X_{\text{Union}}$.
            \item Predict the PSs using the union set: $\hat{Z} = \hat{m}(\X_{\text{Union}})$. 
        \end{enumerate}

        \item For the outcome model $g(z, \X)$ for $z = 0, 1$.
        \begin{enumerate}
            \item Fit an OLS regression of $Y_z$ on $\X_{\text{Union}}$.
            \item Predict the outcomes using the union set: $\hat{Y}_z = \hat{g}(z, \X_{\text{Union}})$. 
        \end{enumerate}
    \end{itemize}

    \item Estimation of the ATE by plugging the predicted values $\hat{m}(\X_{\text{Union}})$ and $\hat{g}(z, \X_{\text{Union}})$ into Equation \ref{equ:dml estimator}.
    
\end{enumerate}
\end{procedurebox}
\noindent The covariate selection steps for the PS and outcome models are identical to those in the estimation procedure using separate covariate sets described in \hyperref[box:doublelasso]{Box 1}. The primary difference is the construction of the union set, $\X_{\text{Union}} = \X_{\text{PS}} \cup \X_{Y_0} \cup \X_{Y_1}$. After constructing the union set, the PS and outcome models are re-estimated using the same union set.

Because the PS and outcome models are re-estimated using the same union set, differential covariate selection no longer occurs. In addition, using the union set increases the likelihood that both models include important covariates needed to remove confounding bias. In particular, the use of the union set allows the DR property to hold even when both models are misspecified based on their respective selected covariate sets. For example, suppose that both selected sets $\X_{\text{PS}}$ and $\X_{Y_z}$ are insufficient to correctly specify their respective models, whereas the union set $\X_{\text{Union}}$ is sufficient to correctly specify at least one model. In this case, the double post-Lasso procedure described in \hyperref[box:doublelasso]{Box 1} fails to remove confounding bias because both models remain misspecified. In contrast, the double post-Lasso procedure using the union set described in \hyperref[box:doublelassounion]{Box 2} provides an additional opportunity for at least one model to be correctly specified, thereby removing confounding bias through the DR property.

\section{Simulation}

\subsection{Data-Generating Model}
In the simulation study, we demonstrate the advantages of using the union set for DR-DML estimation relative to using separate covariate sets for the PS and outcome models. Beyond this primary contribution, the simulation also provides two additional insights. First, by comparing conventional DR estimation with ML-based approaches, we identify conditions under which ML-based covariate selection is beneficial and conditions under which it may be unnecessary or even undesirable. Second, we demonstrate the advantages of post-Lasso over standard Lasso, showing that re-estimating the selected models without penalization can further reduce confounding bias by removing the shrinkage imposed on the estimated coefficients. To address these objectives, we estimated the ATE across a series of simulation scenarios that vary the specification of the PS and outcome models, as well as conditions affecting the performance of Lasso-based covariate selection. 
\begin{figure}[t!] 
\begin{center}
\caption{Causal graph of the data-generating process for treatment $Z$ and outcome $Y$}
\label{fig:dgp}
\includegraphics[width=0.5\textwidth]{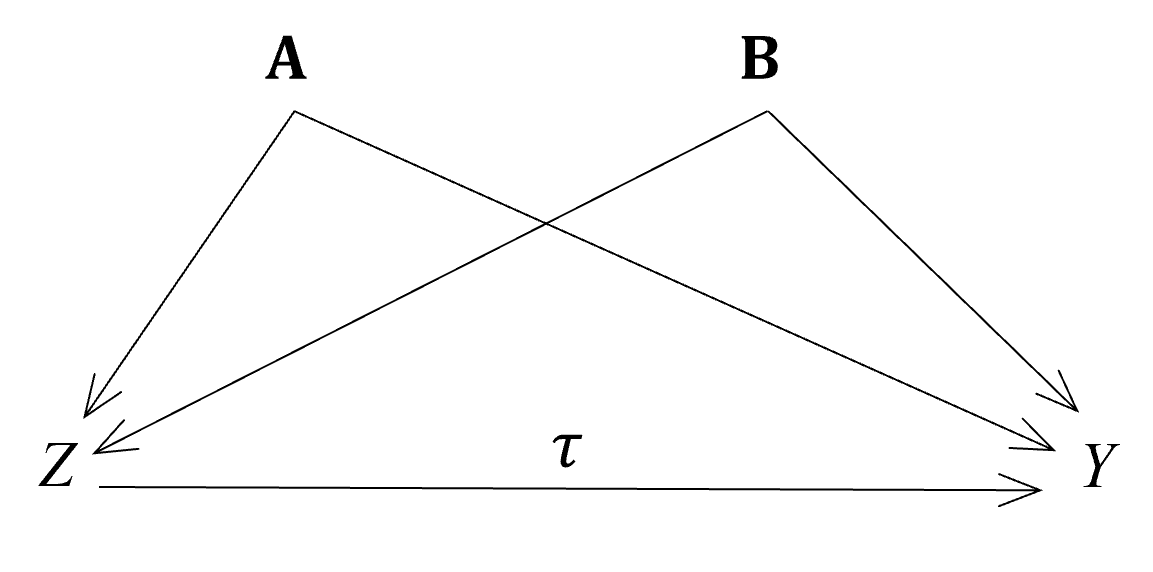}\par
\end{center}
\end{figure}
We generated the data according to the DGP with heterogeneous treatment effects specified in Equation \ref{equ:IRM}, extended to include two independent sets of confounders, as depicted in the causal graph in Figure \ref{fig:dgp}. In the graph, $Y$ denotes the outcome, $Z$ is a binary treatment indicator with $Z = 1$ denoting treatment and $Z = 0$ denoting control, and $\A = (A_k)_{k=1}^p$ and $\B = (B_k)_{k=1}^p$ are two $p$-dimensional sets of observed confounders that jointly satisfy the unconfoundedness assumption. Thus, adjustment for both sets, $\Set = \{ \A, \B \}$, removes the entire confounding bias. The covariates in both confounder sets were independently sampled from a uniform distribution over $[-1, 1]$. The uniform distribution was chosen to avoid extreme PSs and lack of 
overlap. The treatment indicator $Z$ follows a Bernoulli distribution with probability determined by the logistic model $p = \mathbb{P}(Z = 1|\A, \B) = 1/[1 + \text{exp}\{-\boldsymbol{\beta}_Z^{\prime}(\A + \B) \}$, and the outcomes follow the linear model $Y = Z + \boldsymbol{\beta}_Y^{\prime}(\A + \B) + \boldsymbol{\beta}_{YZ}^{\prime}Z(\A + \B) + \varepsilon$, with $\varepsilon \sim N(0, \sigma^2)$. The inclusion of treatment-confounders interactions allows heterogeneous treatment effects. We generated the data so that the confounding bias attributable to each set of confounders is comparable across the PS and outcome models. Allowing the two sets to induce substantially different amounts of confounding bias could itself affect the extent to which each approach reduces bias, introducing an additional source of variation that is beyond the focus of the present study. Three sample-size conditions were considered: $n=500$, $2000$, and $10000$. For the number of covariates, two conditions were considered: $p=50$ and $150$, corresponding to total numbers of covariates of $2p = 100$ and $300$, respectively, across the two confounder sets.

For effective covariate selection, the Lasso relies on the approximate sparsity assumption, under which a relatively small number of covariates have substantial predictive contributions while the remaining covariates have progressively smaller contributions.
Approximate sparsity requires that the ordered absolute values of the coefficients decay sufficiently rapidly. To investigate how the rate of coefficient decay affects the resulting causal effect estimators, the coefficients ($\boldsymbol{\beta}_Z, \boldsymbol{\beta}_Y,$ and $\boldsymbol{\beta}_{YZ}$) were generated as positive values that decay at different rates. Specifically, for each coefficient vector, the $j^{th}$ largest coefficient $\beta_{j}$ was specified to decrease as a function of $j$. For $\beta_{Z,j} \in \boldsymbol{\beta}_Z$ and $\beta_{Y,j} \in \boldsymbol{\beta}_Y$, the coefficients were specified as $\beta_{Z,j}=\beta_{Y,j}=1/j^r,$ with $j=1,\ldots,p$, where $r$ denotes the coefficient decay rate. The coefficients for the treatment-by-confounder interaction, $\beta_{YZ,j} \in \boldsymbol{\beta}_{YZ}$, were specified as $\beta_{YZ,j} = 1/(p+j)^r$ for $j=1,\ldots,p$. Given that the approximate sparsity assumption is satisfied when $r > 0.5$ \citep{chernozhukov2024applied}, two values of the decay rate were considered: $r = 0.51$ and $1$, representing conditions that marginally satisfy, and more strongly satisfy the approximate sparsity condition, respectively. 

\subsection{Estimation of Average Treatment Effect}
Based on the specified data-generating model, we generated 2000 datasets for each simulation condition. For each dataset, the ATE was estimated using the DR-DML estimator in Equation \ref{equ:dml estimator} under four different approaches. The first approach is conventional DR estimation, which estimates the ATE without using the Lasso. Under this approach, the PS model is estimated using logistic regression and the outcome model is estimated using the OLS linear regression. The second approach is doubly robust double Lasso, in which the Lasso is used to estimate both the PS and outcome models. Rather than re-estimating the two models using the covariates selected by the Lasso, the PSs and outcomes are predicted directly from the two fitted Lasso models with shrunken coefficients. The third approach is the doubly robust double post-Lasso procedure described in \hyperref[box:doublelasso]{Box 1}, in which the PS and outcome models are re-estimated using the covariates selected separately from each Lasso model. The fourth approach is the doubly robust double post-Lasso procedure using the union set, described in \hyperref[box:doublelassounion]{Box 2}, in which both models are re-estimated using the union of the covariates selected by the PS and outcome models. For the three Lasso-based approaches, the penalty level (i.e., tuning parameter) was specified using a theoretically derived plug-in value rather than a value selected through cross-validation. The plug-in penalty choice proposed by \citet{belloni2012sparse} is designed to provide theoretical performance guarantees for the Lasso and post-Lasso. This choice is constructed to provide sufficient control over overfitting, thereby reducing the need for cross-validation and additional sample splitting for tuning-parameter selection \citep{chernozhukov2024applied}.\footnote{Comparable theoretically derived plug-in choices for tuning parameters have not yet been established for other ML methods \citep{chernozhukov2024applied}.}

To examine the performance of the four approaches under different model-specification conditions, we considered four scenarios (Scenarios 1–4) representing different combinations of correct and incorrect specification of the PS and outcome models. We define model specification by whether the covariate sets initially included in the respective models contain the covariates needed for correct specification.\footnote{Initial covariate sets are typically chosen based on subject-matter knowledge of the underlying DGP.} Because the Lasso-selected sets may differ from these initial sets, correct specification may not be explicitly defined based on the selected sets. In Scenario 1, both the PS and outcome models include the full covariate set $\Set = \{\A,\B\}$ as their initial covariate sets and are therefore correctly specified. In Scenario 2, the PS model is correctly specified by including $\Set$, whereas the outcome model is misspecified by including only $\B$, that is, omitting $\A$. Scenario 3 represents the reverse case, with the PS model misspecified by including only $\A$ and the outcome model correctly specified by including $\Set$. In Scenario 4, both models are misspecified, with the PS model including only $\A$ and the outcome model including only $\B$.

Note that we intentionally consider these somewhat artificial splits of the initial covariate sets, which may not reflect how researchers would specify the models in practice. The primary purpose of this study is to demonstrate the advantages of using the union set relative to the other approaches across the four model specification scenarios. Thus, a systematic comparison across these scenarios requires holding other simulation conditions constant, including the decay rate governing the approximate sparsity assumption. Alternatively, both the PS and outcome models could begin with the same initial covariate set in every scenario, with the Lasso subsequently selecting different subsets that yield the model specification conditions of interest. However, generating the four scenarios in this manner requires substantially different simulation conditions across scenarios, thereby making it more difficult to attribute differences in estimation performance specifically to the model specification conditions.

In Scenario 1, under the first approach---conventional DR estimation without covariate selection---the conditions for the DR property are satisfied because both the PS and outcome models are correctly specified using $\Set$. Under the second approach---doubly robust double Lasso estimation---the PS and outcome Lasso models separately select covariates from the initial set $\Set$, retaining those with nonzero but shrunken coefficient estimates. Specifically, the PS Lasso model selects $\Set_\text{PS} \subset \Set$, and the outcome Lasso model selects $\Set_Y \subset \Set$. Then, the DR property holds if the PS model is correctly specified using $\Set_\text{PS}$ along with the corresponding shrunken coefficient estimates, or the outcome model is correctly specified using $\Set_Y$ along with the corresponding shrunken coefficient estimates. The third approach---doubly robust double post-Lasso estimation described in \hyperref[box:doublelasso]{Box 1}---uses the same model-specific Lasso selection procedure as the second approach, yielding $\Set_\text{PS} \subset \Set$ for the PS model and $\Set_Y \subset \Set$ for the outcome model. However, rather than using the shrunken Lasso coefficient estimates for predicting PSs and outcomes, the PS and outcome models are re-estimated using conventional regression methods with their respective selected covariate sets to adjust the shrunken coefficient estimates. Thus, the DR property holds if the PS model is correctly specified using $\Set_\text{PS}$, or the outcome model is correctly specified using $\Set_Y$. The fourth approach---doubly robust double post-Lasso estimation with the union set described in \hyperref[box:doublelassounion]{Box 2}---first performs the same separate covariate selection using the PS and outcome Lasso models. It then constructs a union set containing all covariates selected by either model, $\Set_{\text{PS}} \cup \Set_Y$. Both the PS and outcome models are subsequently re-estimated using conventional regression methods with this common union set. Thus, the conditions for the DR property are satisfied if either the PS model or the outcome model is correctly specified using the union set.

\begin{table}[t!]
\caption{Model specification scenarios and conditions for the double robustness property to hold}
\label{tab:covselection}
\vspace{-1.5em}
\begin{center}
\begin{tabular}{lcccc}
\toprule
& Scenario 1 & Scenario 2 & Scenario 3 & Scenario 4 \\
\midrule
\multicolumn{5}{l}{\textbf{Model Specification}} \\
\addlinespace
\multicolumn{5}{l}{\textit{Propensity score model}} \\
Initial set 
& $\Set$ 
& $\Set$ 
& $\A$ 
& $\A$ \\
Specification 
& Correct 
& Correct 
& Incorrect 
& Incorrect \\

\addlinespace

\multicolumn{5}{l}{\textit{Outcome model}} \\
Initial set 
& $\Set$ 
& $\B$ 
& $\Set$ 
& $\B$ \\
Specification 
& Correct 
& Incorrect 
& Correct 
& Incorrect \\

\midrule
\multicolumn{5}{l}{\textbf{Conditions for Double Robustness Property to Hold}} \\
\addlinespace

\multirow{2}{*}{DR}
& \multicolumn{1}{l}{$C_{\text{PS}}(\Set)$ or}
& \multirow{2}{*}{$C_{\text{PS}}(\Set)$}
& \multirow{2}{*}{$C_Y(\Set)$}
& \multirow{2}{*}{---} \\
& \multicolumn{1}{l}{$C_Y(\Set)$}
&
&
& \\

\addlinespace[0.3em]

\multirow{2}{*}{DR-DL}
& \multicolumn{1}{l}{$C_{\text{PS}}(\Set_{\text{PS}})^*$ or}
& \multirow{2}{*}{$C_{\text{PS}}(\Set_{\text{PS}})^*$}
& \multirow{2}{*}{$C_Y(\Set_Y)^*$}
& \multirow{2}{*}{---} \\
& \multicolumn{1}{l}{$C_Y(\Set_Y)^*$}
&
&
& \\

\addlinespace[0.3em]

\multirow{2}{*}{DR-DPL}
& \multicolumn{1}{l}{$C_{\text{PS}}(\Set_{\text{PS}})$ or}
& \multirow{2}{*}{$C_{\text{PS}}(\Set_{\text{PS}})$}
& \multirow{2}{*}{$C_Y(\Set_Y)$}
& \multirow{2}{*}{---} \\
& \multicolumn{1}{l}{$C_Y(\Set_Y)$}
&
&
& \\

\addlinespace[0.3em]

\multirow{2}{*}{DR-DPL(U)}
& \multicolumn{1}{l}{$C_{\text{PS}}
  (\Set_{\text{PS}} \cup \Set_Y)$ or}
& \multicolumn{1}{l}{$C_{\text{PS}}
  (\Set_{\text{PS}} \cup \B_Y)$ or}
& \multicolumn{1}{l}{$C_{\text{PS}}
  (\A_{\text{PS}} \cup \Set_Y)$ or}
& \multicolumn{1}{l}{$C_{\text{PS}}
  (\A_{\text{PS}} \cup \B_Y)$ or} \\

&
\multicolumn{1}{l}{$C_Y
  (\Set_{\text{PS}} \cup \Set_Y)$}
& \multicolumn{1}{l}{$C_Y
  (\Set_{\text{PS}} \cup \B_Y)$}
& \multicolumn{1}{l}{$C_Y
  (\A_{\text{PS}} \cup \Set_Y)$}
& \multicolumn{1}{l}{$C_Y
  (\A_{\text{PS}} \cup \B_Y)$} \\

\bottomrule

\end{tabular}
\end{center}
\vspace{-1em}
\small
\textit{Note.} $\Set=\{\A, \B\}$ denotes the full covariate set. $\Set_{\text{PS}}$ and $\Set_Y$ denote the covariate sets selected from $\Set$ by the propensity score and outcome Lasso models, respectively. $\A_{\text{PS}}$ denotes the subset selected from $\A$ by the propensity score Lasso model, and $\B_Y$ denotes the subset selected from $\B$ by the outcome Lasso model. DR refers to the conventional doubly robust estimation without covariate selection. DR-DL refers to doubly robust double Lasso estimation. DR-DPL refers to doubly robust double post-Lasso estimation. DR-DPL(U) refers to doubly robust double post-Lasso estimation with the union set. $C_{\text{PS}}(\mathcal{X})$ and $C_Y(\mathcal{X})$ indicate that the propensity score and outcome models, respectively, are correctly specified using covariate set $\mathcal{X}$. An asterisk ($*$) denotes correct specification based on both the selected covariates and the corresponding shrunken coefficient estimates. A dash indicates that neither model can be correctly specified using its initial covariate set and, therefore, that the double robustness property does not hold.
\end{table}

Table \ref{tab:covselection} summarizes the model specification scenarios considered in the simulation (upper panel) and the conditions under which the DR property holds for each of the four approaches across these model specification scenarios (lower panel). Table \ref{tab:covselection} shows that in Scenario 4, none of the approaches except the fourth approach can satisfy the conditions required for the DR property to hold. Without using the union set, the PS and outcome models may still jointly remove a substantial portion of the confounding bias induced by $\A$ and $\B$, because each model may eliminate most of the confounding bias attributable to the confounder set it includes. However, because the two models adjust for different sets of confounders, the additional bias amplified by including $\B$ in the outcome model is not addressed by the PS model that includes only $\A$. By contrast, the fourth approach uses the union of the selected covariates to specify both models, allowing each model to incorporate the covariates selected by the other and thereby address bias-inducing misspecification arising from differential covariate selection \citep{steiner2024robust}.

For each simulation condition, we report the proportion of confounding bias remaining, calculated by dividing the bias of each estimator by the original confounding bias, to quantify the extent to which each method removes confounding bias. The simulation results for Scenario 3 in Table \ref{tab:covselection} are reported in the Appendix because Scenarios 2 and 3 can be considered similar with respect to the DR property: in both scenarios, one of the PS or outcome models is correctly specified by its initial covariate set.

\subsection{Results}
\subsubsection{Scenario 1: Initial Sets Correctly Specify Both PS and Outcome Models}

\begin{figure}[t]
\caption{Proportion of confounding bias remaining in average treatment effect estimates for Scenario 1: When initial sets correctly specify both PS and outcome models}
\label{fig:result1}
\vspace{-1.5em}
\begin{center}
\begin{tabular}{c|ccc}
\toprule
Decay & \multirow{2}{*}{$n = 500$} & \multirow{2}{*}{$n = 2000$} & \multirow{2}{*}{$n = 10000$} 
\\ rate & & & \cr
\midrule
\vbox to2cm{\vspace*{0.1cm}\hbox to0.6cm{\hfil $0.51$\hfil}} & \includegraphics[width = 0.28\textwidth]{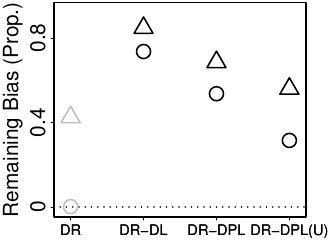} & \includegraphics[width = 0.28\textwidth]{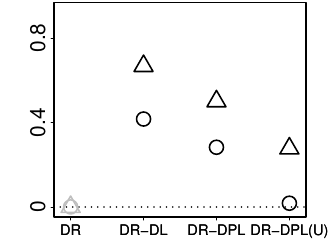} & \includegraphics[width = 0.28\textwidth]{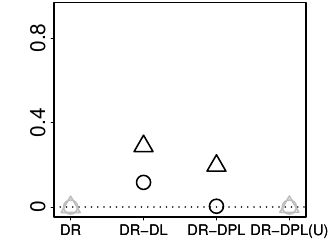} \cr 
\vbox to2cm{\vspace*{0.1cm}\hbox to0.6cm{\hfil $1$\hfil}} & \includegraphics[width = 0.28\textwidth]{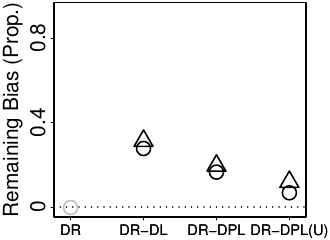} &  \includegraphics[width = 0.28\textwidth]{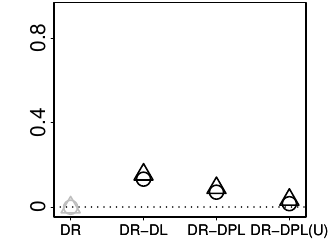} & \includegraphics[width = 0.28\textwidth]{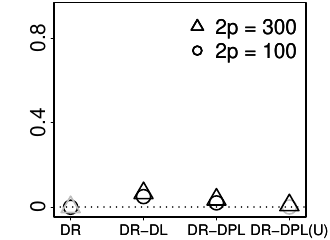} \cr 
\bottomrule
\end{tabular}
\end{center}
\vspace{-1em}
\small
\textit{Note.} $n$ denotes the sample size, and $2p$ denotes the total number of covariates. DR refers to conventional doubly robust estimation without covariate selection. DR-DL refers to doubly robust double Lasso estimation. DR-DPL refers to doubly robust double post-Lasso estimation. DR-DPL(U) refers to doubly robust double post-Lasso estimation using the union set. The horizontal dotted line represents zero remaining bias. Bias with 95\% bootstrap confidence intervals including zero is shown in gray.
\end{figure}

Figure \ref{fig:result1} presents the proportion of confounding bias remaining under Scenario 1, in which the initial covariate sets correctly specify both the PS and outcome models. Overall, conventional DR estimation produces nearly unbiased estimates across most simulation conditions. An exception occurs when the sample size is small relative to the number of covariates. In particular, when $n = 500$ and $2p = 300$, conventional DR estimation yields unstable ATE estimates. Under this condition, the proportion of confounding bias remaining exhibits substantial variability when the decay rate is $0.51$ and becomes markedly larger when the decay rate is $1$, with the value falling outside the range displayed in the figure. 

The performance of the Lasso-based approaches is also less favorable when the sample size is small and the coefficient decay rate is slow. Under these conditions, the Lasso-based estimators generally do not attain the performance of conventional DR estimation and leave a larger proportion of confounding bias remaining. A smaller sample size provides less information for identifying covariates that are important for the PS and outcome models, increasing the likelihood that relevant covariates are omitted during the Lasso selection procedure. This difficulty is further exacerbated under slower coefficient decay, where the predictive signal is distributed across a larger number of covariates rather than concentrated among a relatively small number of strongly predictive covariates. As both the sample size and coefficient decay rate increase, covariate selection becomes more reliable and the relevant predictors become easier to identify, resulting in a smaller proportion of confounding bias remaining and ATE estimates that are increasingly comparable to those obtained from conventional DR estimation.

Among the Lasso-based approaches, the clearest advantage is observed for DR-DPL(U), which consistently exhibits the smallest proportion of confounding bias remaining across simulation conditions. Using the union of the covariate sets selected by the PS and outcome models substantially reduces the remaining bias relative to using separate selected covariate sets, with DR-DPL(U) approaching the performance of conventional DR estimation under many conditions. Even when both models are expected to be approximately correctly specified using their respective selected covariate sets, the union further improves ATE estimation beyond relying on model-specific covariate selection alone. The advantage of using the union is generally more pronounced under conditions with slower coefficient decay, where covariate selection is more challenging. The results also demonstrate the advantage of post-Lasso estimation over standard Lasso. Among the approaches that do not use the union, DR-DPL generally exhibits less remaining confounding bias than DR-DL, indicating that re-estimating the selected models without penalization improves ATE estimation by removing the coefficient shrinkage induced by the Lasso. Accordingly, the overall pattern among the Lasso-based approaches is that DR-DL retains the greatest proportion of confounding bias, DR-DPL reduces this bias through post-Lasso re-estimation, and DR-DPL(U) achieves the greatest reduction by additionally re-estimating both models using the union set.

\begin{figure}[t]
\caption{Proportion of confounding bias remaining in average treatment effect estimates for Scenario 2: When initial sets correctly specify the PS Model but misspecify the outcome Model}
\label{fig:result2}
\vspace{-1.5em}
\begin{center}
\begin{tabular}{c|ccc}
\toprule
Decay & \multirow{2}{*}{$n = 500$} & \multirow{2}{*}{$n = 2000$} & \multirow{2}{*}{$n = 10000$} 
\\ rate & & & \cr
\midrule
\vbox to2cm{\vspace*{0.1cm}\hbox to0.6cm{\hfil $0.51$\hfil}} & \includegraphics[width = 0.28\textwidth]{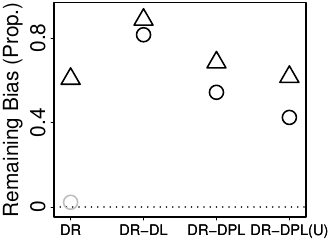} & \includegraphics[width = 0.28\textwidth]{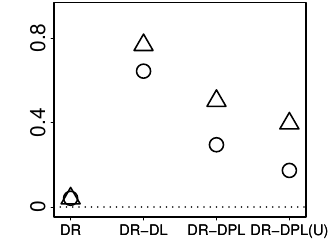} & \includegraphics[width = 0.28\textwidth]{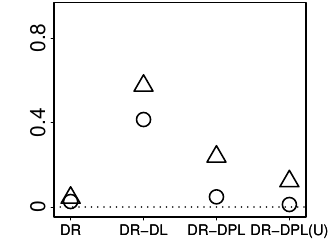} \cr 
\vbox to2cm{\vspace*{0.1cm}\hbox to0.6cm{\hfil $1$\hfil}} & \includegraphics[width = 0.28\textwidth]{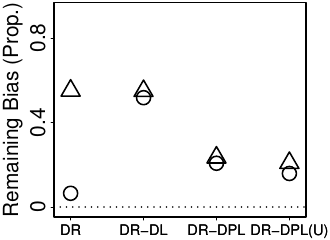} &  \includegraphics[width = 0.28\textwidth]{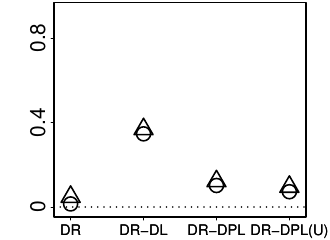} & \includegraphics[width = 0.28\textwidth]{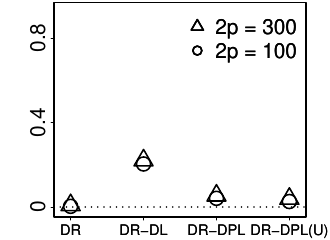} \cr 
\bottomrule
\end{tabular}
\end{center}
\vspace{-1em}
\small
\textit{Note.} $n$ denotes the sample size, and $2p$ denotes the total number of covariates. DR refers to conventional doubly robust estimation without covariate selection. DR-DL refers to doubly robust double Lasso estimation. DR-DPL refers to doubly robust double post-Lasso estimation. DR-DPL(U) refers to doubly robust double post-Lasso estimation using the union set. The horizontal dotted line represents zero remaining bias. Bias with 95\% bootstrap confidence intervals including zero is shown in gray.
\end{figure}

\subsubsection{Scenario 2: Initial Sets Correctly Specify the PS Model but Misspecify the Outcome Model}
Figure \ref{fig:result2} presents the proportion of confounding bias remaining under Scenario 2, in which the PS model is correctly specified by the initial covariate set but the outcome model is misspecified. Overall, the patterns observed under Scenario 1 also emerge under Scenario 2, although the proportion of confounding bias remaining is generally larger. Unlike Scenario 1, in which the DR property can be satisfied through correct specification of either the PS model or the outcome model, the DR property under Scenario 2 depends solely on correct specification of the PS model because the outcome model is misspecified. 

\begin{figure}[t]
\caption{Proportion of confounding bias remaining in average treatment effect estimates for Scenario 4: When initial sets misspecify both PS and outcome models}
\label{fig:result4}
\vspace{-1.5em}
\begin{center}
\begin{tabular}{c|ccc}
\toprule
Decay & \multirow{2}{*}{$n = 500$} & \multirow{2}{*}{$n = 2000$} & \multirow{2}{*}{$n = 10000$} 
\\ rate & & & \cr
\midrule
\vbox to2cm{\vspace*{0.1cm}\hbox to0.6cm{\hfil $0.51$\hfil}} & \includegraphics[width = 0.28\textwidth]{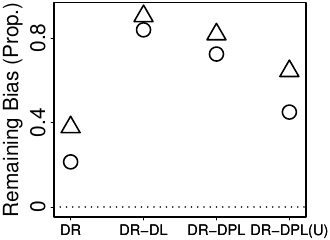} & \includegraphics[width = 0.28\textwidth]{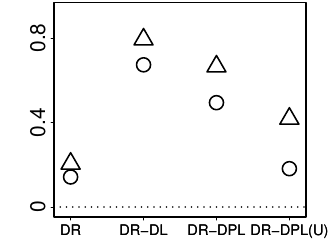} & \includegraphics[width = 0.28\textwidth]{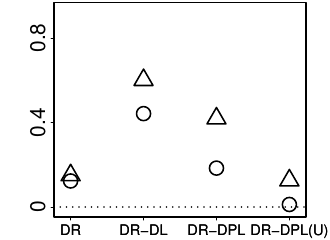} \cr 
\vbox to2cm{\vspace*{0.1cm}\hbox to0.6cm{\hfil $1$\hfil}} & \includegraphics[width = 0.28\textwidth]{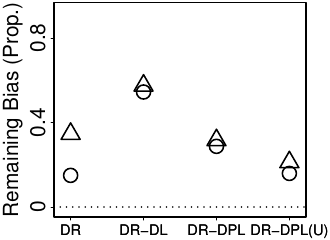} &  \includegraphics[width = 0.28\textwidth]{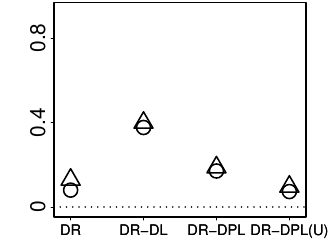} & \includegraphics[width = 0.28\textwidth]{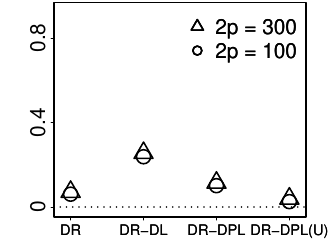} \cr 
\bottomrule
\end{tabular}
\end{center}
\vspace{-1em}
\small
\textit{Note.} $n$ denotes the sample size, and $2p$ denotes the total number of covariates. DR refers to conventional doubly robust estimation without covariate selection. DR-DL refers to doubly robust double Lasso estimation. DR-DPL refers to doubly robust double post-Lasso estimation. DR-DPL(U) refers to doubly robust double post-Lasso estimation using the union set. The horizontal dotted line represents zero remaining bias. Bias with 95\% bootstrap confidence intervals including zero is shown in gray.
\end{figure}

\subsubsection{Scenario 4: Initial Sets Misspecify Both PS and Outcome Models}
Figure \ref{fig:result4} presents the proportion of confounding bias remaining under Scenario 4, in which both the PS and outcome models are misspecified by their respective initial covariate sets. In contrast to the preceding scenarios, all approaches that do not use the union retain non-negligible confounding bias because neither model addresses the bias-inducing misspecification arising from covariates included only in the other model. By contrast, DR-DPL(U) uses the union of the selected covariates to specify both models, allowing each model to incorporate covariates selected by the other and thereby address bias-inducing misspecification arising from differential covariate selection \citep{steiner2024robust}. Accordingly, DR-DPL(U) consistently exhibits substantially less remaining bias than the other approaches, with its advantage becoming more pronounced as the sample size increases. Under some conditions, DR-DPL(U) approaches or slightly outperforms conventional DR estimation.

\section{Discussion}
\subsection{Summary}
High-dimensional data create substantial challenges for causal effect estimation because researchers must identify an appropriate set of covariates from a potentially large number of candidates while also specifying the functional relationships among treatment, outcome, and covariates. ML methods offer useful tools for addressing these challenges by automating covariate selection and flexibly estimating functional forms. DML further facilitates the use of ML for causal inference by mitigating regularization and overfitting bias and thereby permitting $\sqrt{n}$-consistent estimation under appropriate conditions. However, much of the DML literature has emphasized this convergence-rate property arising from Neyman orthogonality, whereas the implications of covariate selection for the DR property possessed by some DML estimators have received comparatively less attention. 

The present study focused on covariate selection as a central component of correct model specification in DR-DML estimation. In particular, we considered two problems that may limit the practical utility of the DR property: differential covariate selection and misspecification of both the PS and outcome models. These problems may be especially relevant in high-dimensional settings because ML methods applied separately to the two models are unlikely to select identical sets of covariates, even when both models begin with the same initial set. Moreover, the selected set for either model may omit covariates necessary for correct model specification. Thus, the use of ML for automated covariate selection does not by itself ensure that the conditions required for the DR property will hold.  

Our primary contribution is to address these two problems arising from incorrect covariate selection by re-estimating both the PS and outcome models using the union of the covariates selected by the two ML models. Moving beyond the limited use of the union in previous studies, we explicitly distinguish the respective roles of DR-DML estimation and the union of selected covariates and combine them to address these two problems. This use of the union is motivated directly by the conditions required for the DR property to hold. re-estimating both models using the same union set eliminates differential covariate selection at the final model-specification stage. At the same time, because the union retains covariates selected as important by either model, it increases the opportunity for at least one nuisance model to include the covariates necessary for correct specification. Thus, rather than simply transferring the conventional use of the union to a different DGP, our approach isolates its distinct role in constructing covariate sets and leverages that role to strengthen the practical utility of the DR property in DR-DML estimation.

The simulation results provide support for our proposed use of the union. Among the Lasso-based approaches, DR-DPL(U) consistently exhibited the smallest proportion of confounding bias remaining. Its advantage was evident even when both the PS and outcome models were approximately correctly specified using their separately selected covariate sets, a setting in which the DR property would already be expected to hold. The benefit of using the union became more pronounced when both models were misspecified by their respective initial covariate sets. Under this condition, approaches using separate selected sets retained non-negligible confounding bias, whereas DR-DPL(U) substantially reduced the remaining bias and, under some conditions, approached or slightly outperformed conventional DR estimation. By using the union, each model incorporates the covariates selected by the other, thereby directly addressing bias-inducing misspecification arising from covariates included only in the other model. The union therefore provides protection not only against differential covariate selection but also against the possibility that neither model is correctly specified by its own selected set.

Beyond this primary contribution, the simulation provides several additional insights into the use of ML for causal inference. First, the results clarify when ML-based covariate selection may and may not be advantageous relative to conventional DR estimation. When an appropriate set of covariates was already available and could be estimated reliably using conventional methods, conventional DR estimation generally performed well. However, conventional DR estimation became unstable when the number of covariates was large relative to the sample size, illustrating the motivation for dimension reduction through ML. At the same time, the Lasso-based procedures could retain substantial confounding bias when the sample size was small relative to the number of candidate covariates. Importantly, these results were obtained under simulation conditions that were relatively favorable to Lasso-based approaches and may be stronger than those typically encountered in practice. In particular, we considered settings in which the approximate sparsity assumption was satisfied and constructed the DGP so that the conditions for the DR property were likely to hold, except in Scenario 4, where both models were misspecified. We also designed the DGP so that the union of the selected covariate sets was likely to retain the covariates needed to satisfy unconfoundedness. Thus, even under conditions favorable to Lasso-based covariate selection, Lasso-based approaches did not uniformly outperform conventional estimation. These findings therefore do not support the automatic use of ML whenever high-dimensional covariates are available. Rather, the value of ML-based estimation depends on the dimensionality of the problem, the available sample size, and the reliability with which the ML procedure can identify the covariates needed for correct specification of the PS and outcome models.

Second, the simulation results highlight the distinction between standard Lasso and post-Lasso estimation. DR-DPL generally retained less confounding bias than DR-DL, indicating that directly using the shrunken coefficients produced by the Lasso can adversely affect causal effect estimation even when Lasso provides useful covariate selection. The improved performance of DR-DPL relative to DR-DL demonstrates the advantage of retaining the covariate-selection function of the Lasso while removing its coefficient shrinkage when estimating causal effects. In this respect, Lasso and post-Lasso should not be treated interchangeably when they are used for causal effect estimation; the distinction between covariate selection and subsequent estimation of the PS and outcome models has meaningful consequences for remaining confounding bias.

Third, the simulation results illustrate the importance of the assumptions underlying the ML procedure itself. For the Lasso, effective covariate selection depends on approximate sparsity. We explicitly varied the coefficient decay rate to represent conditions that marginally and more strongly satisfied approximate sparsity. Lasso-based approaches generally performed less favorably under slower coefficient decay because predictive information was distributed across a larger number of covariates, making it more difficult for the Lasso to retain all covariates needed for correct model specification. As coefficient decay became faster and sample size increased, covariate selection became more reliable and the Lasso-based estimators increasingly approached conventional DR estimation.  These findings emphasize that the performance of ML-based causal estimators depends not merely on the choice to use ML, but also on whether the assumptions and structural conditions that make a particular ML procedure effective are sufficiently plausible for the data at hand.

Finally, although our primary focus was the reduction of confounding bias in finite samples, the proposed union strategy is compatible with the $\sqrt{n}$-consistency that motivates DML estimation. The DR-DML estimator remains based on the same Neyman-orthogonal score; the proposed approach changes how the covariate sets used to estimate the nuisance functions are constructed rather than replacing the underlying causal effect estimator. Thus, re-estimating the PS and outcome models using the union set does not sacrifice the asymptotic motivation for DML while providing an opportunity to improve the removal of confounding bias in finite samples.

\subsection{Limitations}
Several limitations should be considered when interpreting these findings. First, although our simulation extends beyond the DGPs and corresponding DML estimators commonly examined in the DML literature by allowing for heterogeneous rather than constant treatment effects, the DGP itself remains deliberately simple and stylized. In particular, we specified relatively simple functional forms for the PS and outcome models, generated the two sets of confounders independently, and used controlled divisions of the available covariates across simulation scenarios to represent different forms of model specification. These choices were intentional because our primary purpose was to demonstrate the distinct advantages of using the union set and to isolate the consequences of covariate selection from other complexities of the DGP. Nevertheless, such settings do not fully reflect the complexity of real-world high-dimensional data, which may involve violations of unconfoundedness and approximate sparsity, nonlinear relationships and complex dependence structures among covariates. Future research should therefore evaluate the proposed approach under more realistic and complex DGPs, particularly settings in which both covariate selection and estimation of the functional forms of the PS and outcome models are challenging. For example, when the true DGP involves interactions among confounders or higher-order polynomial effects, greater attention must be given to how \textit{predictors} are initially constructed, because the candidate predictor set may need to include nonlinear transformations and interaction terms derived from the original covariates. Unless such relevant terms are explicitly included among the candidate predictors, Lasso-based selection may fail to adequately approximate the underlying PS or outcome regression function. Future studies should therefore consider richer predictor representations or ML methods that can more flexibly capture complex functional relationships.

Second, the present study considers only the Lasso as the ML method. The Lasso provides an important starting point because of its foundational role in the development of high-dimensional causal inference and because its explicit covariate-selection mechanism makes it possible to clearly distinguish the effects of covariate selection, coefficient shrinkage, and the union set. Nevertheless, the generality of the proposed strategy should be investigated using a broader range of ML methods. In particular, future work could consider tree-based ensemble methods such as random forests and boosting, which can implicitly select or prioritize predictors through their fitting procedures. Extending the proposed approach to such methods will require defining how information about selected or important predictors is combined across the PS and outcome models. It will also be important to examine the learner-specific conditions that play a role analogous to approximate sparsity for the Lasso and to determine how violations of those conditions affect both covariate selection and the performance of the resulting DR-DML estimator. The current results already demonstrate that the effectiveness of Lasso-based procedures depends materially on the degree of approximate sparsity, suggesting that the benefits of the union should be evaluated jointly with the assumptions underlying the ML method used to construct it. 

Taken together, the findings suggest that covariate selection should be treated as an integral component of DR-DML estimation rather than as an independent preprocessing step conducted separately for the PS and outcome models. The union set provides a simple way to integrate the information learned by the two models, address differential covariate selection, and increase the opportunity for the DR property to hold when model-specific selected sets are insufficient. More broadly, our findings emphasize that the advantages of ML for causal inference depend not only on the predictive performance of the learning algorithm, but also on the assumptions underlying the ML procedure and on how the information learned from the data is incorporated into the causal estimator.

\bibliography{ma_ref_all}
\end{document}